\documentclass[9pt,twocolumn,twoside]{opticajnl}
\journal{opticajournal} % Choose journal (ao,jocn,josaa,josab,ol,optica,pr)

\setboolean{shortarticle}{false}
\usepackage{lineno}
\usepackage{mathtools}
\usepackage{relsize}
\title{20 GHz fiber-integrated femtosecond pulse and supercontinuum generation with a resonant electro-optic frequency comb}

\author[1,2,*]{Pooja Sekhar}
\author[1,2]{Connor Fredrick}
\author[2,4]{David R. Carlson}
\author[2,4]{Zachary Newman}
\author[1,2,3,\dag]{Scott A. Diddams}

\affil[1]{Department of Physics, University of Colorado Boulder, 440 UCB, Boulder, Colorado 80309, USA}
\affil[2]{Time and Frequency Division, National Institute of Standards and Technology, 325 Broadway, Boulder, Colorado 80305, USA}
\affil[3]{Department of Electrical, Computer and Energy Engineering, University of Colorado Boulder, Colorado 80309, USA}
\affil[4]{Octave Photonics, 325 W South Boulder Rd, Louisville, Colorado 80027, USA}

\affil[*]{Corresponding author: pooja.sekhar@colorado.edu}

\affil[$\dag$]{email: scott.diddams@colorado.edu}

\begin{abstract}
Frequency combs with mode spacing in the range of 10  to 20 gigahertz (GHz) are critical for increasingly important applications such as astronomical spectrograph calibration, high-speed dual-comb spectroscopy, and low-noise microwave generation. While electro-optic modulators and microresonators can provide narrowband comb sources at this repetition rate, a significant remaining challenge is a means to produce pulses with sufficient peak power to initiate nonlinear supercontinuum generation spanning hundreds of terahertz (THz) as required for self-referencing in these applications. %While integrated photonics can provide narrowband comb sources at this repetition rate, there is presently no integrated solution capable of producing hundreds of THz at $\sim 10$ GHz as required for these applications. In this context, the biggest challenge remains a means to produce pulses with sufficient peak power to initiate nonlinear supercontinuum generation spanning hundreds of terahertz (THz). 
Here, we provide a simple, robust, and universal solution to this problem using off-the-shelf polarization-maintaining (PM) amplification and nonlinear fiber components. This fiber-integrated approach for nonlinear temporal compression and supercontinuum generation is demonstrated with a resonant electro-optic frequency comb at 1550 nm. We show how to readily achieve pulses shorter than 60 fs at a repetition rate of 20 GHz and with peak powers in excess of 2 kW. The same technique can be applied to picosecond pulses at 10 GHz to demonstrate temporal compression by a factor of $9\times$ yielding 50 fs pulses with peak power of 5.5 kW. %Employing an improved fiber design, we also predict generating a sub-60 fs pulse from the 2.2 ps wide temporal output of a cascaded electro-optic comb. 
These compressed pulses enable flat supercontinuum generation spanning more than 600 nm after propagation through multi-segment dispersion-tailored anomalous-dispersion highly nonlinear fiber (HNLF) or tantala waveguides. The same 10 GHz source can readily achieve an octave-spanning spectrum for self-referencing in dispersion-engineered silicon nitride waveguides. This simple all-fiber approach to nonlinear spectral broadening fills a critical gap for transforming any narrowband 10 to 20 GHz frequency comb into a broadband spectrum for a wide range of applications that benefit from the high pulse rate and require access to the individual comb modes.
\end{abstract}

\setboolean{displaycopyright}{true}

\begin{document}

\maketitle

\section{Introduction}

There is growing interest in broadband, low-noise optical frequency combs at repetition rates in the range of 10 to 30 gigahertz (GHz) for a wide range of applications such as astronomical spectrograph calibration \cite{Braje2008AstronomicalCombs,Metcalf2019StellarComb, Suh2019SearchingAstrocomb}, high-speed precision spectroscopy \cite{Kowligy2020Mid-infraredGHz, Carlson2020BroadbandMeasurements, Martin-Mateos2015DualSpectroscopy, Suh2016MicroresonatorSpectroscopy, Dutt2018On-chipSpectroscopy}, low-noise microwave generation \cite{Lucas2020Ultralow-noiseOscillator, Yang2021Dispersive-waveSources, Fortier2011GenerationDivision}, and optical arbitrary waveform generation \cite{Cundiff2010OpticalGeneration}. Our work is particularly motivated by the need for simple and robust approaches to 10+ GHz comb generation for astronomical spectrograph calibration. Initial efforts to generate such high repetition rate combs for astronomy relied on filtering a subset of modes from mode-locked lasers (MLL) \cite{Quinlan2010ACalibration, Ycas2012DemonstrationComb, SteinmetzT2008LaserObservations, Kirchner2009GenerationMultiplication, Zou2016BroadbandSpectrographs, Probst2020ACombs}. However, studies have shown that this approach can lead to calibration errors due to the amplification of unwanted modes through four-wave mixing \cite{Ycas2012DemonstrationComb,Probst2013NonlinearCombs}. As a result, recent works have directly realized astrocombs with large mode spacings using electro-optic modulation (EOM or EO) \cite{Yi2016DemonstrationAstronomy, Metcalf2019StellarComb}, microcombs \cite{Suh2019SearchingAstrocomb, Obrzud2019AAstrocomb} and cascaded four-wave mixing \cite{ChavezBoggio2018WavelengthAstro-comb, Myslivets2012GenerationDispersion}. 

\begin{figure*}[htb]
\centering\includegraphics[width=18cm]{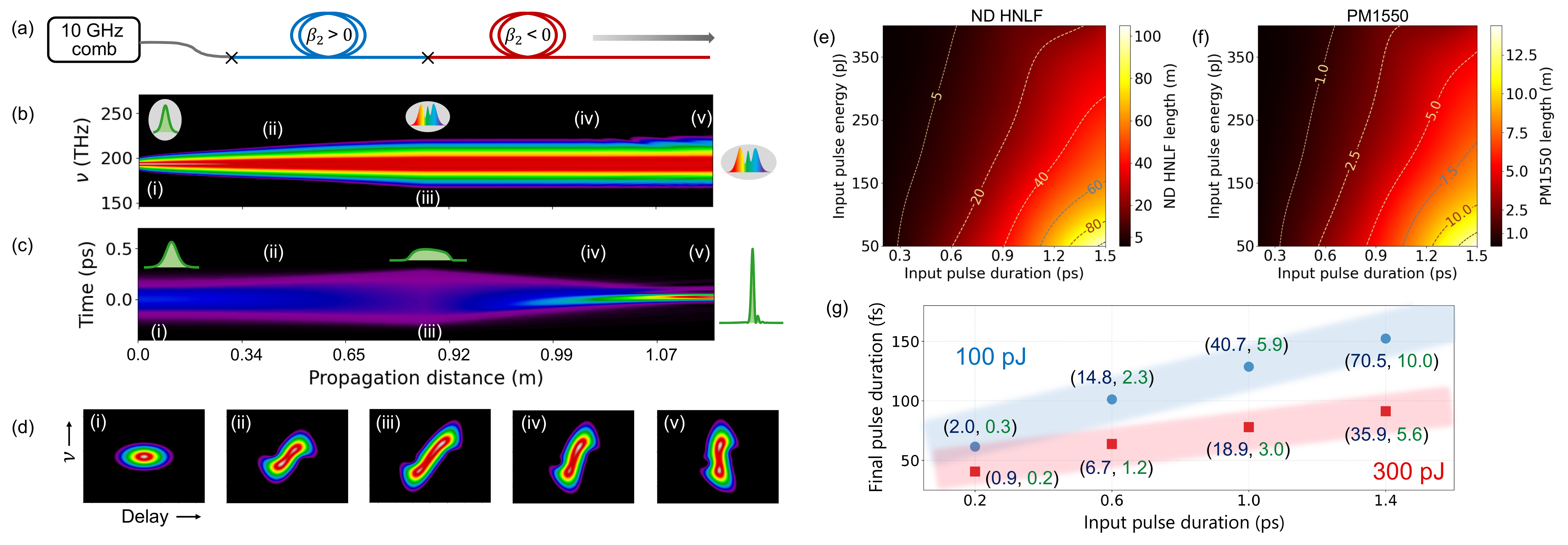}
\caption{All-fiber approach for nonlinear temporal compression. (a)~Schematic of the all-fiber temporal compressor. The 10 GHz comb pulses are amplified and sent through ND HNLF ($\beta_{2} > 0$) for spectral broadening and then through an appropriate length of PM1550 with anomalous dispersion ($\beta_{2} < 0$) at 1550 nm to produce sub-50 fs pulses. (b)~Simulated spectral evolution of an ideal Gaussian pulse with full width at half maximum (FWHM) pulse duration of 250 fs and pulse energy of 250 pJ propagated through the temporal compression stage consisting of 0.9 m ND HNLF (D = -2.6 ps/(nm$\cdot$ km)) and 21 cm of PM1550. (c)~Simulated temporal evolution of the 250 pJ, 250 fs input Gaussian pulse through the all-fiber temporal compressor. The compressed pulse is 49 fs. (d)~Spectrograms of the pulse as it propagates through the all-fiber temporal compressor. The images correspond to the propagation distances marked in parts (b) and (c). (e)~Optimal ND HNLF length (in meters) for temporal compression of a Gaussian pulse with initial duration and energy in the range of 0.2 ps - 1.5 ps and 50 pJ - 400 pJ. (f)~Corresponding PM1550 length (in meters) required to compensate the normal chirp due to the ND HNLF. (g)~Pulse durations achievable through simulation of the all-fiber temporal compressor with various input pulse durations and two different pulse energies of 100 pJ and 300 pJ. The parentheses list the required ND HNLF and PM1550 fiber lengths (in meters).}
\label{scheme}
\end{figure*}

Electro-optic modulation of a continuous wave (CW) laser to generate frequency combs with GHz mode spacing has the additional benefits of simplicity, robustness, reliability, and flexibility in independently tuning the operating wavelength as well as the mode spacing \cite{Torres-Company2014OpticalPhotonics, Ishizawa2011GenerationDiode, Zhang2019BroadbandResonator, Hu2022High-efficiencyGenerators}. However, generating EO frequency combs with sufficient bandwidth typically requires a cascaded series of phase modulators (PMs) driven with multiple Watts of microwave power. For example, in order to generate a comb with a bandwidth of 10 nm, each cascaded modulator is typically driven by radio frequency signals with average powers in excess of 3 W. As a result, the entire comb generation process is quite costly and the components require active cooling for long-term operation. A simpler and more cost-effective approach employs the novel technique of resonant EO modulation, as first implemented by Kourogi \textit{et. al}. \cite{KourogiM1993Wide-spanMeasurement}. Utilizing highly efficient, resonant EO comb generators, frequency combs with tens of GHz mode spacing and bandwidths of more than 50 nm can be generated using about ten times lower RF power as compared to cascaded EO modulators \cite{Carlson2018UltrafastControl}. %Another promising feature of the resonant EO modulation technique is its integrated implementation in a thin-film lithium niobate nanophotonic platform \cite{Zhang2019BroadbandResonator, Hu2022High-efficiencyGenerators}. 
We leverage this comb generation approach using a commercially available package, consisting of a fiber-coupled waveguide EO phase modulator within a resonant Fabry-P\'erot cavity \cite{Saitoh1995AGenerator}, to generate frequency combs at 1.55 $\mu$m with repetition rates in excess of 10 GHz. % Loncar integrated : interesting opportunity, nanophotonic etched waveguide, cite coupling one: extra efficiency, maybe mention this where we specify 25 dB loss 

While electro-optic frequency combs and microcombs directly provide mode spacings in the range of tens of GHz, a remaining challenge for these sources lies in matching this large mode spacing with the spectral coverage (hundreds of THz) required, for example, to span the entire bandwidth of an astronomical spectrograph. EOM combs \cite{Metcalf2013High-powerGenerator, Metcalf201930Visible} and microcombs \cite{Lamb2018Optical-FrequencySupercontinuum, Suh2019SearchingAstrocomb, Herr2014TemporalMicroresonators} only give rise to a few hundred femtosecond (fs) pulses in the time domain whereas traditional mode-locked lasers \cite{Li2010Diode-pumpedPower, Bartels2007SpectrallyLaser, Klenner2014HighLaser, Bartels2008PassivelyLaser, Chen20133GHzWavelength, Klenner2014GigahertzLaser} can directly yield sub-100 fs pulses as required for low noise supercontinuum generation \cite{Corwin2003FundamentalFiber, Dudley2002NumericalFibers}. Despite many efforts in this direction, producing sufficient pulse peak power to initiate nonlinear processes for broadband supercontinuum generation still remains a big challenge for bulk as well as chip-integrated frequency combs at tens of GHz repetition rates \cite{Yu2022FemtosecondLens}. %Earlier efforts showed that sub-100 fs input pulse duration is required to obtain good signal-to-noise (SNR) for the comb modes in the broadband supercontinuum generated in anomalous dispersion media \cite{Corwin2003FundamentalFiber, Kubota1999AnalysesInteraction, Dudley2002NumericalFibers}. 
To this end, previous studies have generally explored nonlinear temporal compression from a few hundred femtosecond (fs) to $\sim$ 50 fs in two ways: (i) soliton self-compression in fibers \cite{Tamura1999FemtosecondFiber, Tamura200154-fsCompressor,Pekarek2011Self-referenceableLaser}, nanophotonic waveguides \cite{Oliver2021Soliton-effectChip}, and microresonators \cite{Obrzud2017TemporalPulses}, (ii) self-phase modulation in multi-pass gas cells (in the microjoule pulse energy regime) \cite{Viotti2022Multi-passPulses, Balla2020PostcompressionRegime} or highly nonlinear fibers (HNLFs), where the chirp on the resulting spectrally-broadened pulses is compensated using diffraction gratings \cite{Lesko2020Fullym, Hoghooghi2022BroadbandComb, Carlson2018UltrafastControl, Metcalf201930Visible, Hartl2009FullyComb}, pulse shapers \cite{Cole2016Octave-spanningMultiplication, Beha2017ElectronicLight, Wu2013Supercontinuum-basedGeneration_Weiner} or anomalous dispersion fibers \cite{Obrzud2018BroadbandComb, Obrzud2019VisibleGeneration, Zhang2020Sub-100m, Ataie2013GenerationSource}. 
%Nevertheless, a simple and all-fiber approach that can compress 20 GHz comb pulses with near-to-complete compensation of nonlinear spectral phase has not been demonstrated to the best of our knowledge. 

In this work, we demonstrate a simple and all-PM-fiber approach to temporal compression of 20 GHz comb pulses from a resonant EO comb generator to 57 fs. Careful compensation of the nonlinear spectral phase results in pulses with less than 5$\%$ of the total energy in the pedestals. Using the same technique, we achieve pulses as short as 50 fs at 10 GHz repetition rate and demonstrate temporal compression by a factor of $9\times$ for picosecond pulses at 10 GHz. Extending our fiber design, we also predict generating a sub-60 fs pulse from the 2.2 ps wide temporal output of a cascaded electro-optic comb. The nonlinear temporal compression is followed by a flat broadband supercontinuum generation in carefully designed multi-segment dispersion-tailored fibers or nanophotonic waveguides. The entire nonlinear system of temporal compression and supercontinuum generation has been modeled and the simulation results show excellent agreement with the experiment. Our approach to nonlinear spectral generation with off-the-shelf telecom components achieves simplicity, robustness, and long-term stability that can be applied to transform any narrowband frequency combs at these repetition rates into a broadband spectrum. 
\section{All-fiber approach for temporal compression to femtosecond pulses}
Our all-fiber approach is based on self-phase modulation (SPM) in a polarization-maintaining normal dispersion highly nonlinear fiber (ND HNLF) and the subsequent dispersion compensation in an appropriate length of anomalous dispersion fiber. The main advantage of this technique over soliton self-compression is the lower noise amplification associated with the inhibition of modulation instability in the normal dispersion regime. Fig. \ref{scheme}(a) illustrates our approach. The 10 GHz comb pulses are amplified and sent through ND HNLF for spectral broadening. As a pulse with high peak power ($P_{0}$) propagates through HNLF with length $L$, an intensity-dependent nonlinear phase shift is imparted to the pulse. The maximum value of this phase shift is given by $\gamma P_{0}L$ at the peak, where $\gamma$ is the nonlinear coefficient of the fiber. As a result, the pulse acquires a normal chirp, i.e. the leading edge of the pulse is red-shifted in frequency and the trailing edge is blue-shifted. In the presence of normal group-velocity dispersion (GVD), the red components travel faster than the blue components, leading to additional chirp on the pulse. 
%However, an analytical expression for spectral broadening becomes more involved with the interplay between SPM and GVD and demands numerical analysis. 
We study the spectral broadening process via numerical modeling of the pulse propagation by solving the generalized nonlinear Schrodinger equation (GNLSE) as given below through the split-step Fourier method in PyNLO \cite{Agrawal2013NonlinearOptics, PyNLO}, \begin{equation} \begin{multlined}
\frac{\partial E(z, t)}{\partial z}=i \mathlarger{\sum}_{k \geq 2} \frac{i^{k} \beta_{k}}{k !} \frac{\partial^{k} E}{\partial t^{k}}+i \gamma\left(1+\frac{i}{\omega_{0}} \frac{\partial}{\partial t}\right) \\ \times\left[E ( z , t )  \left(\int_{-\infty}^{t} R\left(t^{\prime}\right)\left|E\left(z, t-t^{\prime}\right)\right|^{2} d t^{\prime}+i \Gamma_{R}(z, t)\right)\right].
\end{multlined} \label{eq1} \end{equation} 
Here $E(z,t)$ is the complex pulse envelope, $\beta_{k}$ refer to the second and higher orders of fiber dispersion, the second term on the right-hand side of equation \ref{eq1} corresponds to the Kerr nonlinear effects including self-steepening and the last term incorporates Raman effects.

Fig. \ref{scheme}(b) and (c) show the simulated spectral and temporal evolution of a Gaussian pulse with full width at half maximum (FWHM) of 250 fs and 250 pJ energy (average power of 2.5 W at 10 GHz) as it propagates through the temporal compression fiber stage comprised of 0.9 m ND HNLF and 21 cm of PM1550. Here, we are using the parameters of commercial (OFS) ND HNLF with dispersion values of -2.6 ps/(nm$\cdot$km), D\textsubscript{slope} of 0.026 ps/(nm$^{2}\cdot$km), and nonlinearity parameter of 10.5 W$^{-1}$km$^{-1}$ at the pump wavelength of 1550 nm. In simulations, the pulse propagates through ND HNLF until the broadened spectrum supports the shortest possible pulse (sub-100 fs). However, longer HNLF adds higher order dispersion terms to the spectral phase of the pulse that cannot be compensated by the quadratic anomalous dispersion of polarization-maintaining 1550 nm fiber (PM1550). Taking into account both these counteracting effects, the ND HNLF length is fixed at 0.9 m such that the broadened bandwidth supports a short, sub-100 fs pulse while minimizing the higher order spectral phase. The pulse spectrogram at different positions through ND HNLF is shown in Fig. \ref{scheme}(d), which clearly illustrates the normal chirp imparted to the pulse, as well as the spectral and temporal broadening.

In order to obtain close to transform-limited pulse duration, the normal chirp has to be compensated. This is usually done with a grating compressor to provide the required anomalous chirp \cite{Carlson2018UltrafastControl, Lesko2020Fullym, Hoghooghi2022BroadbandComb}. Here, we use an all-fiber approach in which an appropriate length of PM1550 with anomalous dispersion at 1550 nm replaces the grating compressor. The nonlinearity of PM1550 is quite low compared to HNLF. As a result, further spectral broadening is insignificant, and the dispersion compensation is employed to get near transform-limited pulse durations. Fig. \ref{scheme}(c) illustrates the temporal evolution of the given input pulse in the all-fiber temporal compressor to give a compressed pulse duration of 49 fs with 4.4 kW peak power. The output pulse spectrogram in Fig. \ref{scheme}(d) shows that most of the normal chirp imparted to the pulse from ND HNLF has been compensated on propagation through PM1550.

\begin{figure}[h!]
\centering\includegraphics[width=9.2cm]{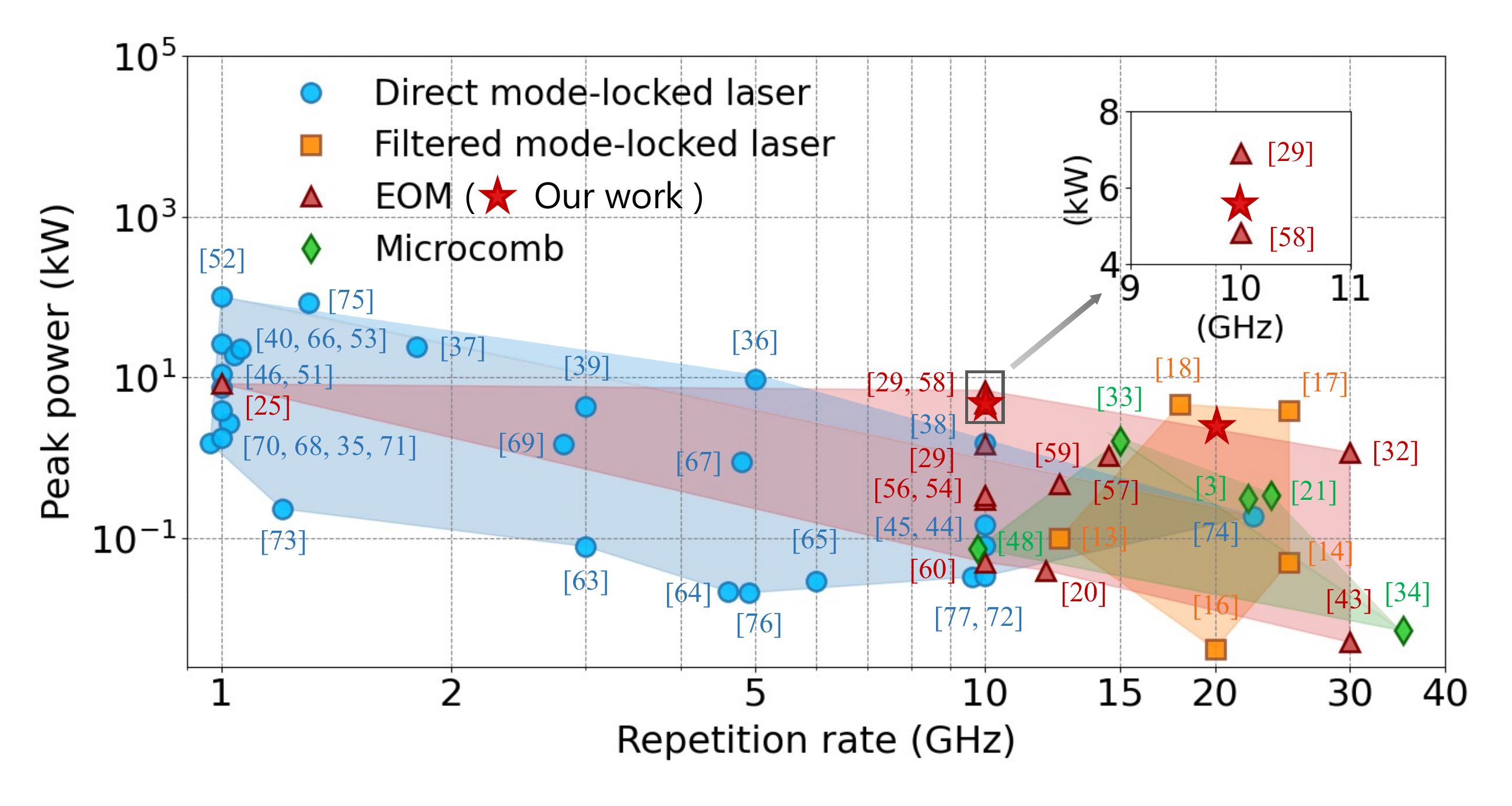}
\caption{Overview of the peak powers obtained through state-of-the-art nonlinear temporal compression of GHz frequency combs. The inset figure shows a zoomed-in portion of the plot as marked by the box at 10 GHz repetition rate. The maximum peak powers achieved in this work at repetition rates of 10 GHz and 20 GHz are labeled as red stars. %We demonstrate one of the highest peak powers for frequency combs with tens of GHz repetition rates. (Have to add references on the plot) Increase plot size
}
\label{sum}
\end{figure}

We generalize this design for various comb generation approaches by performing simulations to estimate the lengths of fiber required to compress a given input pulse at $\sim$10 GHz repetition rates to sub-150 fs. Fig. \ref{scheme}(e) shows the approximate length of ND HNLF and Fig. \ref{scheme}(f) shows the corresponding PM1550 length required for temporal compression of a given input pulse with duration ranging from 200 fs to 1.5 ps and energy varying from 50 pJ to 400 pJ. These values correspond to an average power of 0.5 W to  4 W at 10 GHz repetition rate. Using this combination of ND HNLF and PM1550, Fig. \ref{scheme}(g) provides the achievable compressed pulse duration for given input pulse characteristics. With this technique, it is possible to compress a 1.4 ps input pulse to 150 fs at pulse energies as low as 100 pJ. Even though this gives an estimate of the achievable compressed pulse duration, for the best prediction of the actual compressed pulse duration it is necessary for the input pulse in the simulations to resemble the experimentally recorded one. Fig. \ref{scheme}(g) also illustrates the applicability of this simple technique to a wide range of input pulse durations from hundreds of fs to ps and energies typical of various techniques of GHz comb generation. As we show below, this all-fiber nonlinear temporal compression approach has been applied to a resonant electro-optic comb generator at microwave frequencies of 10 and 20 GHz to obtain pulse durations between 50 fs and 60 fs. The peak powers of the compressed pulses achieved in this work, labeled as red stars in Fig.~\ref{sum} are comparable to state-of-the-art nonlinear temporal compression approaches for GHz repetition rates \cite{Chen20123Laser, Endo2012Kerr-lensRate, Endo20136-GHzSpectroscopy, Klenner2013ASignal, Pekarek2012FemtosecondGHz, Schratwieser2012HighlyLaser, Yamazoe2010Palm-top-sizeMirror, Pekarek2010Diode-pumpedKW, Byun2010CompactLasers, Sobon201110Laser, Yang20121.2-GHzMirror, Zhuang2013High-powerLocking, Kobayashi20101.3-GHzSystem, Choudhary2012Diode-pumpedRate, Martinez2011Multi-gigahertzNanotubes}. A few important features differentiate our work from the previously reported high peak powers at these repetition rates. The average power at 10 to 20 GHz repetition rates in our work is $\sim$ 4$\times$ lower as compared to the mode-filtered cases where the comb is amplified to average powers of 12 W to 15 W to get compressed pulse durations of 130 fs to 140 fs \cite{Zou2016BroadbandSpectrographs, Probst2020ACombs}. The peak power of $\sim$ 6.9 kW inferred from \cite{Carlson2018UltrafastControl} involves a second stage of supercontinuum generation in silicon nitride waveguide and the subsequent dispersion compensation using silica glass in free space in addition to the initial stage of spectral broadening in highly nonlinear fiber and temporal compression using diffraction gratings. % grating, more free space components 

\section{Demonstration in a Resonant Electro-optic Comb}

\begin{figure*}[h!]
\centering\includegraphics[width=17cm]{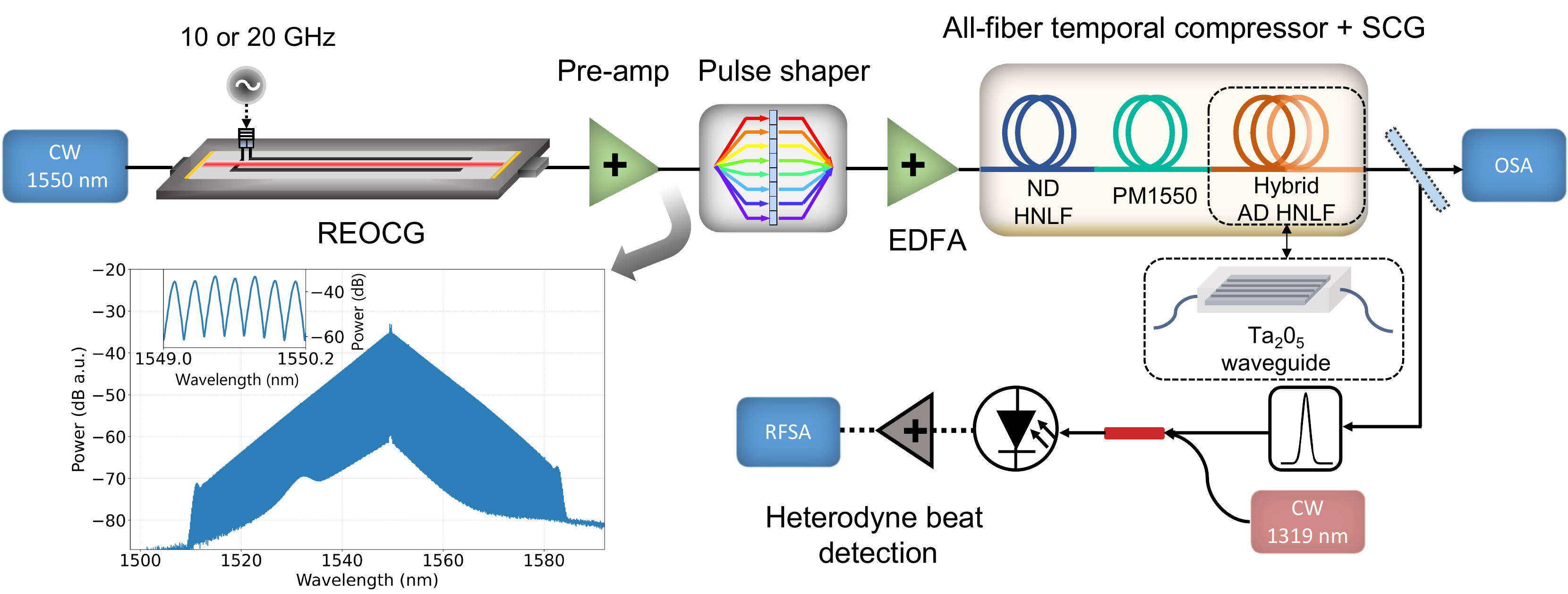}
\caption{Experimental setup for nonlinear temporal compression and supercontinuum generation of the resonant electro-optic frequency comb. The blue trace shows the output of the 20 GHz comb generator and the inset shows the 20 GHz comb modes. The compressed pulses were sent to anomalous dispersion highly nonlinear fibers or nanophotonic waveguides for supercontinuum generation. The light reflected from the beam splitter at the output of the spectral broadening stage is used to measure the heterodyne beat between the filtered supercontinuum and a continuous wave (CW) laser. REOCG: resonant electro-optic comb generator, EDFA: erbium-doped fiber amplifier,  ND HNLF: normal dispersion highly nonlinear fiber, PM1550: polarization-maintaining single-mode fiber at 1550 nm, AD HNLF: anomalous dispersion highly nonlinear fiber, Ta$_{2}$O$_{5}$: tantalum pentoxide, SCG: supercontinuum generator, OSA: optical spectrum analyzer, RFSA: radio frequency spectrum analyzer. }    
\label{setup}
\end{figure*}
\subsection{All-fiber temporal compression to sub-60 fs}

\begin{figure*}[h!]
\centering\includegraphics[width=18cm]{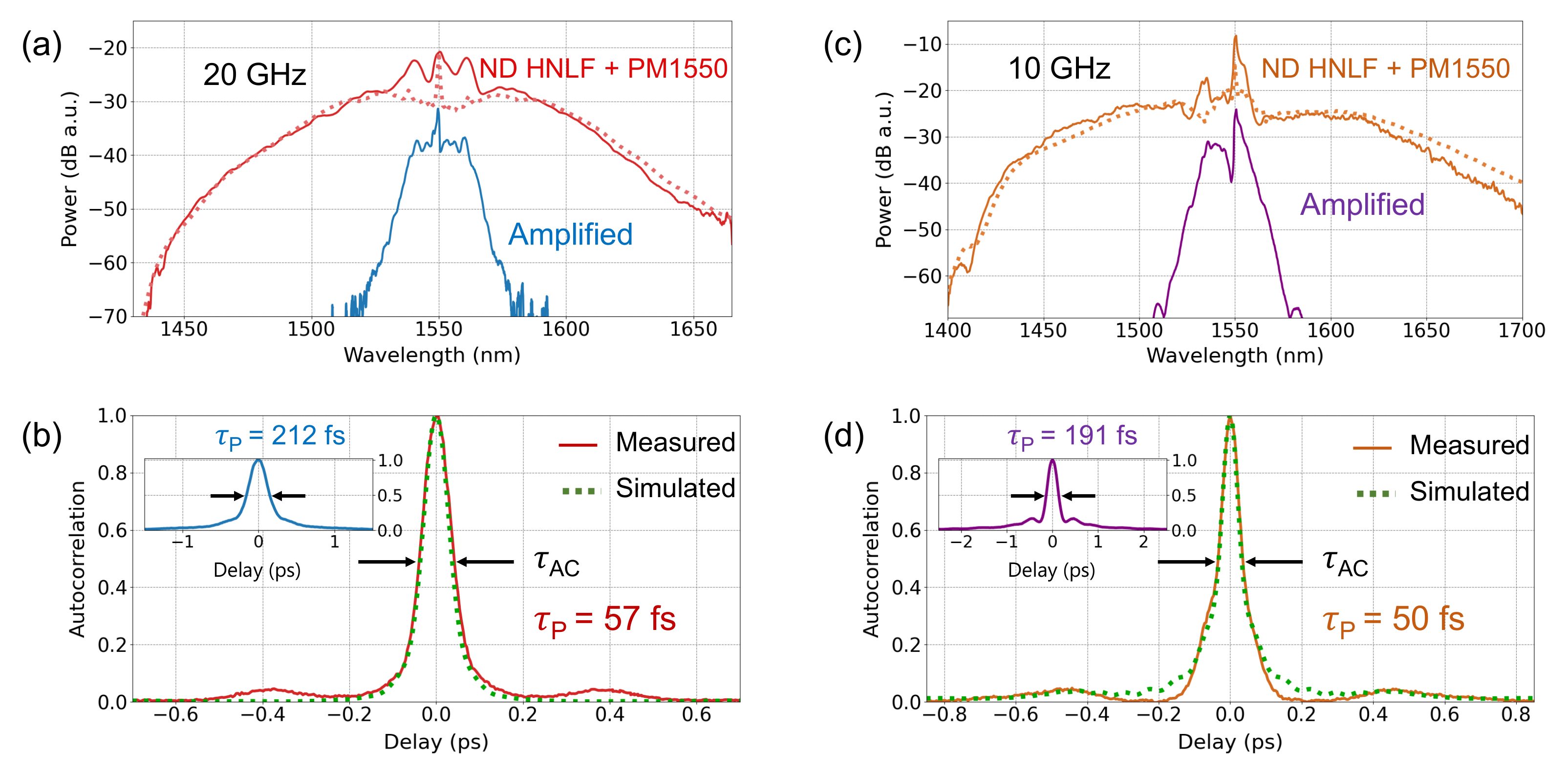}
\caption{(a) Optical spectra recorded after amplification of the 20 GHz comb pulses to 3 W (blue trace) and propagation through a temporal compression stage consisting of 1.2 m of ND HNLF and 22 cm of PM1550 (solid red trace). The dotted red trace gives the simulated spectrum. The traces are offset by 10 dB for clarity. (b) Measured autocorrelation after amplification (blue trace) and temporal compression (red trace) of the 20 GHz comb. The dotted green trace shows the calculated autocorrelation of the compressed pulse obtained from the NLSE simulation. The pulse duration ($\tau_{P}$) is calculated from the autocorrelation width ($\tau_{AC}$) considering a deconvolution factor of 1.41. The measured FWHM of the compressed pulse (57 fs) is very close to the simulated value of 52 fs (dotted green trace). c) Optical spectra after amplification of the 10 GHz comb pulses (purple trace) and propagation through the same fiber temporal compressor (orange trace). The dotted orange trace gives the simulated spectrum. The traces are offset by 15 dB for clarity. (d) Corresponding autocorrelation traces recorded for the 10 GHz comb. The measured FWHM of the compressed pulse at 10 GHz is 50 fs while the simulation (dotted green trace) gives 45 fs.}
\label{comp}
\end{figure*}

Our electro-optic comb generation employs a fiber-integrated waveguide phase modulator in a resonant Fabry-P\'erot cavity (model: WTEC-01) \cite{Saitoh1995AGenerator}. The so-called resonant electro-optic comb generator (REOCG) has the advantages of highly efficient resonance modulation and built-in RF phase noise filtering \cite{Kim2017Cavity-inducedComb}. Fig. \ref{setup} shows the schematic of our experimental setup. A cavity-stabilized continuous wave laser at 1550 nm is used to seed the comb generator. A dielectric resonant oscillator (DRO Synergy model: KDFLOD1000-8) phase-locked to a 10 MHz maser signal is frequency doubled to 20 GHz and then amplified to an average power of 1 W. The REOCG is driven by this 1 W microwave signal at 20 GHz that is a multiple of the Fabry-P\'erot cavity's free spectral range (FSR = 2.5 GHz). The optical spectrum of the resonant EOM comb has a double-sided exponentially decaying shape as shown in Fig. \ref{setup}. More details about the REOCG can be found in the Supplement. The comb pulses are sent through a programmable pulse shaper for dispersion compensation and to apply the required group delay across one side of the spectrum in order to overlap the two interleaved pulse trains while maintaining the entire optical system in fiber-coupled components. The pulse shaper is not critical for achieving close to transform-limited pulses since it can be replaced by an appropriate length of single-mode fiber for nearly quadratic dispersion compensation. The 20 GHz pulse train is then amplified to 3 W average power using an erbium-doped fiber amplifier (EDFA). The intensity autocorrelation width of the amplified pulses after nonlinear broadening in erbium-gain fiber is measured to be $\sim$ 300 fs (inset in Fig. \ref{comp}(b)). Assuming a Gaussian pulse shape with a deconvolution factor of 1.41, the resultant pulse duration is 212 fs which is close to the band-limited pulse duration of 195 fs. This deconvolution factor has been verified in all cases by calculating the autocorrelation pulse duration from the band-limited pulse.

Although the REOCG generates pulses as short as 212 fs after amplification, previous studies have shown that pulse durations greater than 100 fs adversely affect the bandwidth as well as the coherence of the supercontinuum generated in anomalous dispersion fibers \cite{Corwin2003FundamentalFiber, Dudley2002NumericalFibers}. In order to overcome this issue, we compress this pulse to less than 60 fs using an all-fiber approach as described in the previous section. Fig. \ref{scheme}(g) shows that ${\sim}2$ m of normal-dispersion (ND) highly nonlinear fiber (HNLF) broadens an ideal Gaussian pulse of 200 fs and 100 pJ to a spectral bandwidth that supports a 50 fs pulse. Assuming an average power of 2.95 W (pulse energy of 148 pJ at 20 GHz), simulation results show that 1.2 m of ND HNLF is required for the first stage of spectral broadening through self-phase modulation (SPM) as depicted by the dotted red trace in Fig. \ref{comp}(a). To maintain close agreement with the experiment, we anomalously chirp the Fourier transform-limited pulse of the amplified 20 GHz comb spectrum and use it as input to the simulations. On propagating the spectrally broadened pulse through 22 cm of PM1550 to overcome the normal chirp, modeling predicts a compressed pulse duration of 52 fs (Fig. \ref{comp}(b)). Experimentally, we were able to measure a broadened spectrum (solid red trace in Fig. \ref{comp}(a)) and compressed pulse with full width at half maximum (FWHM) of 57 fs after propagating the amplified pulses through 1.2 m of ND HNLF and 22 cm of PM1550, in exact agreement with the model. Fig \ref{comp}(b) shows the autocorrelation traces at the output of PM1550 in our setup (red solid curve) as well as through simulations (dotted green curve) and again demonstrates excellent agreement between the two. The average output power from the PM1550 fiber is measured to be 2.8 W. This results in a 57 fs pulse with a peak power of $\sim$ 2.4 kW.%, which is close to the largest recorded peak powers for ultrashort pulses at 20 GHz repetition rates as shown in Fig. \ref{sum}. 

In order to demonstrate the robustness of our temporal compression scheme in tunable comb systems, we changed the repetition rate of our resonant electro-optic comb to 10 GHz and employed the same setup as illustrated in Fig. \ref{setup}. Here, the pulse shaper is used to remove half of the optical spectrum to generate a single 10 GHz pulse train (more information is provided in the Supplement). The 10 GHz pulses are amplified to 3 W using an EDFA and the resultant intensity autocorrelation pulse duration is measured to be 270 fs, which corresponds to a pulse duration of 191 fs assuming a deconvolution factor of 1.41 (inset in Fig. \ref{comp}(d)). On propagating these amplified pulses through the same fiber compression stage consisting of 1.2 m long ND HNLF (D = -2.6 ps/(nm$\cdot$km)) and 22 cm PM1550, the pulse duration is reduced to 50 fs. The corresponding measured autocorrelation (orange trace) shows good agreement with the simulated result (dotted green trace in Fig. \ref{comp}(d)). The Fourier transform-limited pulse duration of the spectrum recorded after the compression fiber stage is $\sim$ 30 fs. The compressed pulse is slightly chirped as opposed to the 20 GHz case since the fiber compressor design is not optimized for the 10 GHz system.    

\begin{figure}[h!]
\centering\includegraphics[width=9cm]{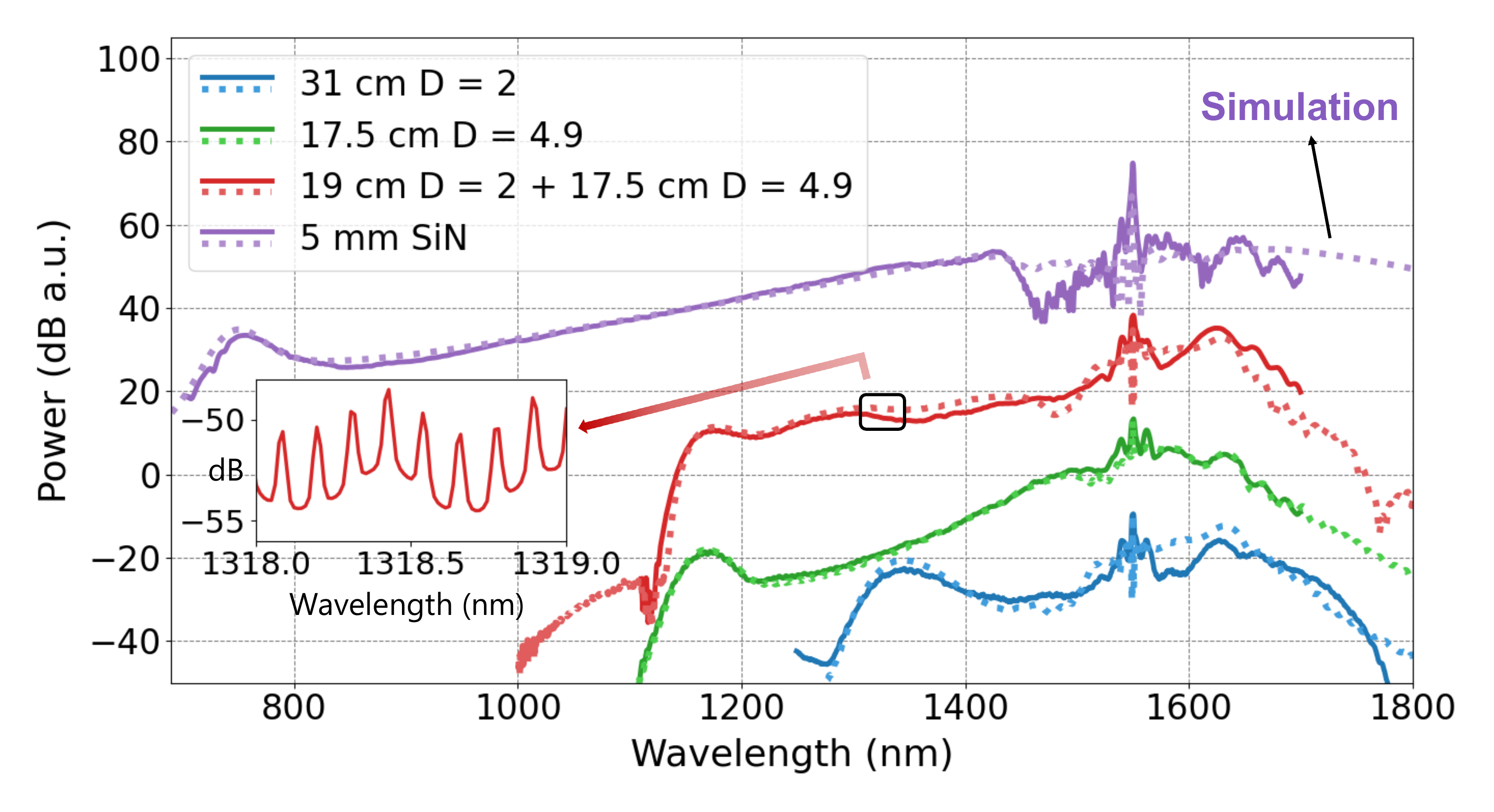}
\caption{Supercontinuum spectra generated with the compressed pulses shown in figure~\ref{comp}. The blue, green, and red traces are from the 57 fs driving pulse at 20 GHz in different lengths of AD HNLF with dispersion values of D = 2.2 ps/nm.km (blue), D = 5.4 ps/nm.km (green) and hybrid (red). The purple trace corresponds to an octave-spanning supercontinuum obtained in a 5 mm long SiN waveguide from the 50 fs driving pulse at 10 GHz. Experimental and simulation results are shown as solid and dotted lines respectively. The inset shows the 20 GHz comb modes around 1319 nm of the supercontinuum generated from hybrid AD HNLF. The blue, green, and red traces are vertically offset by 20 dB, and the purple trace by 30 dB for clarity.}
\label{scg}
\end{figure}
\subsection{Supercontinuum generation in anomalous dispersion HNLF and SiN waveguide}
The compressed pulse at 20 GHz is then passed through anomalous dispersion (AD) HNLF for supercontinuum generation. We studied the supercontinuum generated in two AD HNLFs with different dispersion characteristics independently and observed that a flat broadband spectrum is generated on propagation through multi-segment AD HNLF at the point of soliton fission. Fig. \ref{scg} shows the corresponding spectra obtained experimentally (solid curves) as well as through modeling (dotted curves). The blue trace corresponds to the supercontinuum generated using 31 cm long HNLF with lower anomalous dispersion at the pump wavelength (D = 2.2 ps/(nm$\cdot$km), and D\textsubscript{slope} = 0.026 ps/(nm\textsuperscript{2}$\cdot$km). The green trace indicates the supercontinuum obtained in 17.5 cm long HNLF with higher AD (D = 5.4 ps/(nm$\cdot$km), and D\textsubscript{slope} = 0.028 ps/(nm\textsuperscript{2}$\cdot$km)). The HNLF with lower AD generates a dispersive wave (DW) around 1345 nm (blue trace) whereas the one with higher AD can produce DW at 1156 nm, wavelengths further away from the pump (green trace).% way to determine HOD of HNLF
The total splicing loss through the combined temporal compression and supercontinuum generation fiber system is measured to be $\sim$ 1.5 dB. This loss is attributed to the splicing between PM1550 and smaller core HNLF and has been included in the input pulse energy to the respective fibers in the simulation. Since the dispersion parameters of HNLF given by the manufacturer are only valid over a small wavelength band of $\sim$ 100 nm around 1550 nm, the fiber dispersion parameters in simulations were tuned to match the location of the dispersive wave in the broad supercontinuum observed experimentally. This tuning is performed after initially fixing the input pulse duration and energy to the measured values. The corresponding dispersion parameters used in simulations are D = 2 ps/(nm$\cdot$km), D\textsubscript{slope} = 0.034 ps/(nm\textsuperscript{2}$\cdot$km) and D = 4.9 ps/(nm$\cdot$km), D\textsubscript{slope} = 0.023 ps/(nm\textsuperscript{2}$\cdot$km). This approach gives an efficient way to determine the dispersion parameters of the commercially available HNLF over a broad spectral range.

A smooth supercontinuum spanning 1150 nm to 1700 nm (red trace in Fig. \ref{scg}) is generated in a hybrid HNLF consisting of 19 cm long D = 2 ps/(nm$\cdot$km) and 12 cm long D = 4.9 ps/(nm$\cdot$km). The length of the first lower dispersion AD HNLF is fixed before the point of soliton fission and then the pulse train is propagated through an appropriate length of the second AD HNLF to obtain a flat broader spectrum. The zoomed-in trace of the resolved 20 GHz comb modes of the supercontinuum at 1319 nm measured at 0.02 nm resolution on an optical spectrum analyzer is shown in the inset of Fig. \ref{scg}. The total average power measured in free space after the multi-segment AD HNLF is $\sim$ 2.5 W, which corresponds to more than 20 $\mu$W/mode around the dispersive wave at 1150 nm. 

In order to use these broadband spectra for high-impact metrology applications, we need a self-referenced frequency comb which in turn requires an octave-spanning spectrum. We show that the temporally compressed 50 fs, 10 GHz comb source developed in this work can readily generate an octave-spanning spectrum (purple trace in Fig. \ref{scg}) in a dispersion-engineered silicon nitride (SiN) waveguide. The waveguide fabricated by the open access foundry Ligentec is 5 mm long with a cross-section of 800 nm $\times$ 2500 nm and is silica clad. The average power incident on the waveguide is 2.8 W and the measured total insertion loss is $\sim$ 4.3 dB. Considering this reduction in input pulse energy as well as the calculated frequency-dependent effective index and nonlinearity parameter of the waveguide via finite difference mode solver, the simulated spectrum (dotted purple trace in Fig. \ref{scg}) closely agrees with the recorded one.  

\begin{figure*}[h!]
\centering\includegraphics[width=18cm]{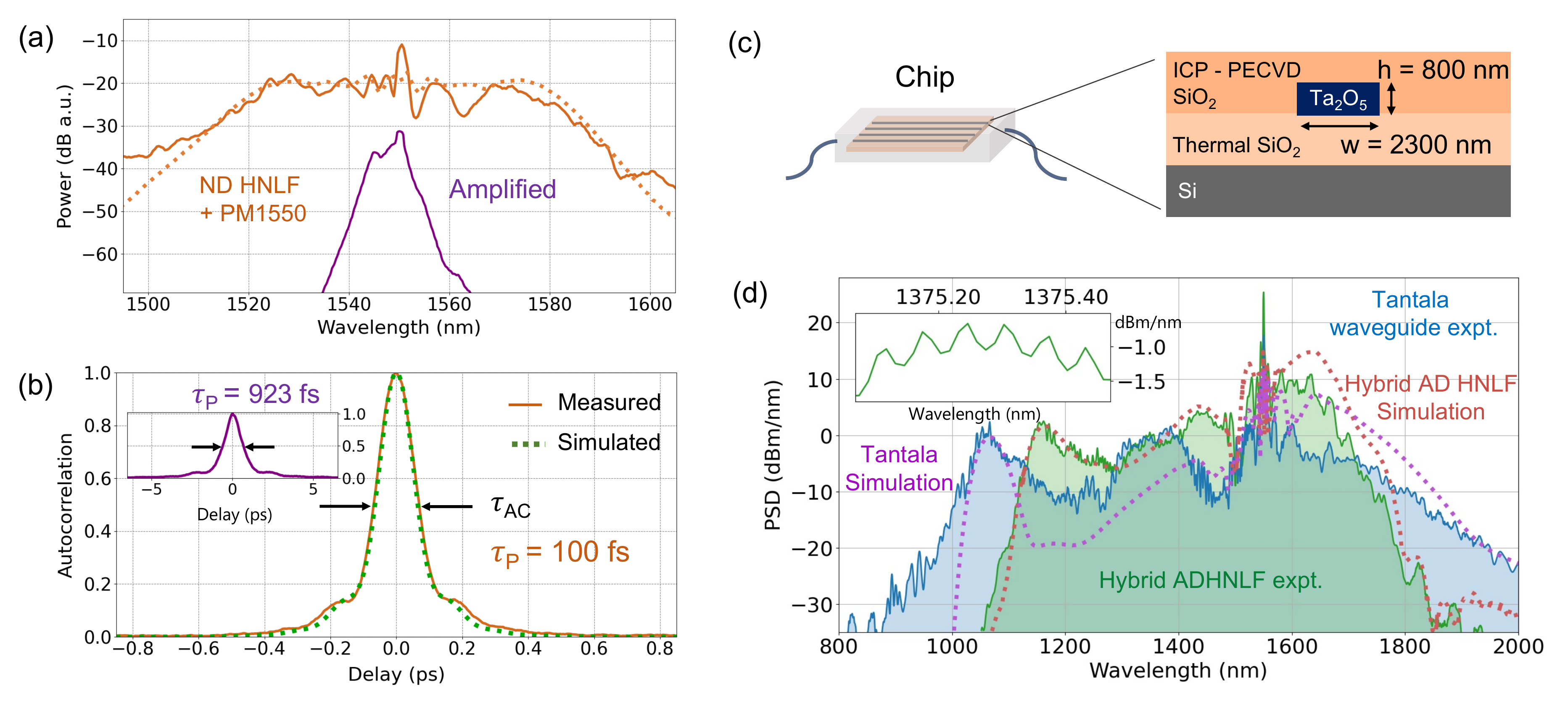}
\caption{(a) Optical spectra after amplification of the spectrally compressed, 920 fs 10 GHz comb pulses to 3 W (purple trace) and propagation through a temporal compression stage consisting of 8.4 m of ND HNLF and 2 m of PM1550 (orange trace). The dotted orange trace shows the simulated spectrum. The traces are offset by 18 dB for clarity. (b) Measured autocorrelation after amplification (purple trace) and temporal compression (orange trace) of the 10 GHz comb. The measured pulse duration ($\tau_{P}$ = 100 fs) is calculated from the autocorrelation width ($\tau_{AC}$) considering a deconvolution factor of 1.41. The dotted green trace shows the calculated autocorrelation of the compressed pulse (duration of 93 fs) obtained from the NLSE simulation. (c) Fiber-connectorized tantala (Ta$_{2}$O$_{5}$) chip and its waveguide cross-section. (d) Supercontinuum spectra generated from the compressed 100 fs 10 GHz pulses in a multi-segment AD HNLF (green trace) consisting of 23 cm of D = 2 ps/(nm$\cdot$km) and 8 cm of D = 5.4 ps/(nm$\cdot$km), and in the tantala waveguide (blue trace). The dotted traces represent simulation results in the multi-segment AD HNLF (red) and tantala waveguide (purple). The inset figure shows the 10 GHz comb modes around 1375 nm of the supercontinuum obtained using the hybrid AD HNLF.}
\label{parvi}
\end{figure*}

\subsection{Extending this approach to picosecond input pulses}

Since electro-optic comb generators generally give picosecond pulses in the time domain, we designed and implemented our all-PM fiber temporal compression technique for longer picosecond input pulses. Initially, the spectral bandwidth of our 10 GHz comb is reduced using the pulse shaper such that the intensity autocorrelation after amplification to 3 W gives 923 fs pulse on deconvolution (inset in Fig. \ref{parvi}(b)). Then, we propagate the amplified pulses through 8.4 m of ND HNLF (D = -2.6 ps/(nm$\cdot$km)) and 2 m of PM1550 to achieve a compressed pulse duration of 100 fs as shown by the orange autocorrelation trace in Fig. \ref{parvi}(b). Assuming chirped transform-limited pulses of the amplified comb spectrum as the input, modeling results on the propagation of the amplified pulses through 8.4 m of ND HNLF and 1.82 m of PM1550 gives a compressed pulse duration of 93 fs, which is in close agreement with the experiment. %Here, the slight disagreement between the simulation and experiment in terms of PM1550 length as compared to the previous cases can be attributed to the fact that the experimental input pulse is not fully bandwidth-limited since the length of PM1550 after amplification and before temporal compression fiber stage was not optimized after the initial bandwidth reduction. 

The resulting 100 fs pulse is then sent through a different multi-segment dispersion-tailored AD HNLF for supercontinuum generation. Fig. \ref{parvi}(d) shows the supercontinuum (blue curve) generated in hybrid AD HNLF consisting of 23 cm long D = 2 ps/(nm$\cdot$km) fiber and 8 cm long D = 5.6 ps/(nm$\cdot$km) fiber, and the corresponding modeled spectrum is depicted in the red dotted trace. These fiber dispersion parameters and the nonlinearity parameter value of 10.5 W\textsuperscript{-1}km\textsuperscript{-1} are given by OFS. The simulation results demonstrate better agreement with the experiment on tuning the HNLF dispersion parameters to be D = 1.3 ps/(nm$\cdot$km), D\textsubscript{slope} = 0.03 ps/(nm\textsuperscript{2}$\cdot$km) and D = 5.8 ps/(nm$\cdot$km), D\textsubscript{slope} = 0.027 ps/(nm\textsuperscript{2}$\cdot$km) respectively for the given input pulse. The lengths of the fibers were fixed so that soliton fission happens at the end of the multi-segment fiber stage to produce a broad flat spectrum from 1160 nm to 1660 nm as shown in the green trace of Fig \ref{parvi}(d). The 10 GHz comb modes around 1375 nm are depicted in the inset of Fig. \ref{parvi}(d). 

Finally, we demonstrate 10 GHz supercontinuum generated from this 100 fs pulse in a specially designed fiber-packaged tantalum pentoxide (Ta$_{2}$O$_{5}$) or tantala waveguide module fabricated by Octave Photonics. The tantala waveguide is 19 mm long with a cross-section of 800 nm $\times$ 2300 nm (Fig. \ref{parvi}(c)). The waveguides have silica cladding and are fabricated on a silicon wafer. The average power incident on the waveguide is $\sim$ 1.7 W and the total insertion loss is measured to be 5.4 dB. We observed a flat supercontinuum extending from 1050 nm to 1700 nm in wavelength (blue trace in Fig. \ref{parvi}(d)). The simulation result in tantala waveguide given by the purple dotted trace in Fig. \ref{parvi}(d) shows reasonable agreement with the experimentally measured spectrum. We estimated the nonlinearity parameter of Ta$_{2}$O$_{5}$ at 1550 nm to be 1.2 W$^{-1}$km$^{-1}$ and calculated the frequency-dependent effective index of the waveguide via finite difference fully vectorial mode solver in our nonlinear pulse modeling. 

\section{Conclusions and Future Work}

We presented a simple, all-PM-fiber approach for efficient nonlinear temporal compression to femtosecond pulses and supercontinuum generation from 10+ GHz frequency combs. Using this scheme, we generated ultrashort pulses of duration 50 fs and 57 fs with peak powers of 5.5 kW and 2.4 kW respectively at 10 GHz and 20 GHz repetition rates. These pulses originated from a resonant fiber-integrated waveguide-type electro-optic comb generator. We have also demonstrated a flat supercontinuum spanning over 500 nm with more than 20 $\mu$W/mode from these compressed pulses in a multi-segment dispersion-tailored AD HNLF. The compressed pulses enabled the generation of an octave-spanning spectrum in a SiN waveguide, which in turn showed the potential to stabilize our 10 GHz comb through self-referencing. By employing this pulse compression scheme, we achieved a high temporal compression ratio of $9\times$ for picosecond input pulses at 10 GHz and generated a supercontinuum that spans 1050 nm to  1700 nm in tantala waveguides. The simulation results on nonlinear pulse propagation have shown excellent agreement with the experiment throughout. 
%Our good understanding on the numerical modelling of the current system can be leveraged to generate an octave-spanning spectrum in a different dispersion-tailored fiber \cite{Cole2016Octave-spanningMultiplication} or in nanophotonic waveguides \cite{Carlson2018UltrafastControl}. 

The simplicity and robustness of our nonlinear spectral broadening technique with off-the-shelf telecom components make it compatible with other techniques for comb generation and well-suited for applications such as astronomical spectrograph calibration and fast acquisition of precise spectroscopic data. Our 10+ GHz comb system also shows promise for generating visible astrocombs via sum frequency generation in periodically poled lithium niobate waveguides \cite{Wu2022UltravioletPoling} and mid-infrared combs for high-speed spectroscopy via intra-pulse difference frequency generation in nonlinear crystals \cite{Lind2020Mid-InfraredOptics, Hoghooghi2022BroadbandComb}. Future work will include the study of the coherence and RIN of the supercontinuum, with a focus on understanding the effect of the intrinsic electro-optic cavity filtering on broadband microwave thermal noise.  

\begin{backmatter}
\bmsection{Funding} 
National Science Foundation (AST 2009982); Jet Propulsion Laboratory (RSA 1671354); National Institute of Standards and Technology (NIST on a Chip). %Content in the funding section will be generated entirely from details submitted to Prism. Authors may add placeholder text in the manuscript to assess length, but any text added to this section in the manuscript will be replaced during production and will display official funder names along with any grant numbers provided. If additional details about a funder are required, they may be added to the Acknowledgments, even if this duplicates information in the funding section. See the example below in Acknowledgements.

\bmsection{Acknowledgments}
The authors thank Tsung-Han Wu for valuable help with fiber splicing and Sida Xing for helpful discussions on nonlinear optics modeling. The authors thank Ryan Cole, Kristina F. Chang, and Daniel H. Slichter for their valuable comments and discussions. We further thank Stephanie Leifer for her early contributions to this work and Michael Geiselmann of Ligentec for providing the SiN waveguides. Certain equipment, instruments, software, or materials, commercial or non-commercial, are identified in this paper in order to specify the experimental procedure adequately. Such identification is not intended
to imply recommendation or endorsement of any product or service by NIST, nor is it intended to imply that the materials or equipment
identified are necessarily the best available for the purpose. %The section title should not follow the numbering scheme of the body of the paper. Additional information crediting individuals who contributed to the work being reported, clarifying who received funding from a particular source, or other information that does not fit the criteria for the funding block may also be included; for example, ``K. Flockhart thanks the National Science Foundation for help identifying collaborators for this work.''

\bmsection{Disclosures} 
David Carlson and Zachary Newman are owners and employees of Octave Photonics.

\bmsection{Data availability} Data underlying the results presented in this paper are not publicly available at this time but may be obtained from the authors upon reasonable request.

\bmsection{Supplemental document}
See Supplement 1 for supporting content.

\end{backmatter}

\bibliography{optica-journal-template, optica-journal-template-2}

\bibliographyfullrefs{optica-journal-template}

\ifthenelse{\equal{\journalref}{aop}}{%
\section*{Author Biographies}
\begingroup
\setlength\intextsep{0pt}
\begin{minipage}[t][6.3cm][t]{1.0\textwidth} % Adjust height [6.3cm] as required for separation of bio photos.
  \begin{wrapfigure}{L}{0.25\textwidth}
    \includegraphics[width=0.25\textwidth]{john_smith.eps}
  \end{wrapfigure}
  \noindent
  {\bfseries John Smith} received his BSc (Mathematics) in 2000 from The University of Maryland. His research interests include lasers and optics.
\end{minipage}
\begin{minipage}{1.0\textwidth}
  \begin{wrapfigure}{L}{0.25\textwidth}
    \includegraphics[width=0.25\textwidth]{alice_smith.eps}
  \end{wrapfigure}
  \noindent
  {\bfseries Alice Smith} also received her BSc (Mathematics) in 2000 from The University of Maryland. Her research interests also include lasers and optics.
\end{minipage}
\endgroup
}{}

\end{document}

\maketitle

\section{Resonant Electro-optic Comb Generator}

\begin{figure}[htb]
\centering\includegraphics[width=13cm]{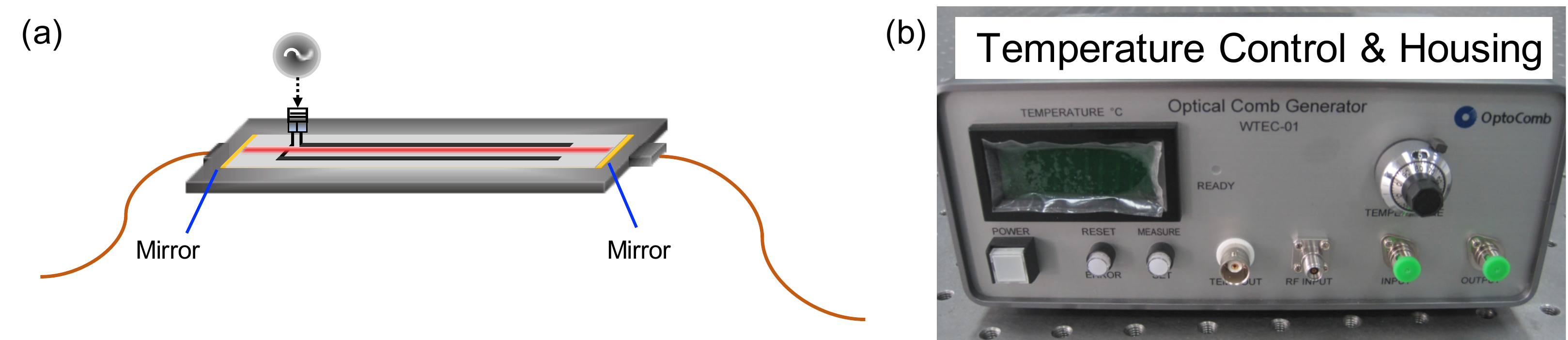}
\caption{(a) Schematic of the resonant electro-optic frequency comb generator (REOCG) that consists of a fiber-integrated waveguide phase modulator with high reflection films on each facet. (b) The commercially available REOCG is kept inside a metal housing.}
\label{scheme}
\end{figure}

We employ a compact Fabry-P\'erot electro-optic frequency comb generator originally developed by Kourogi et. al \cite{KourogiM1993Wide-spanMeasurement}. It consists of a fiber-integrated waveguide phase modulator in a resonant Fabry-P\'erot cavity  that was fabricated by coating high-reflection films on the facets of a lithium niobate waveguide (Fig. \ref{scheme}(a)). The finesse and free spectral range (FSR) of the optical resonator are 58 and 2.5 GHz, respectively \cite{Saitoh1995AGenerator}, which allows us to generate frequency combs with repetition rates at integer multiples of the cavity FSR. This commercially available resonant electro-optic comb generator (REOCG, Optocomb model: WTEC-01 \cite{Saitoh1995AGenerator}) is kept inside a metal housing as shown in Fig. \ref{scheme}(b). The REOCG has a built-in temperature control unit and the temperature of the cavity can be easily tuned to reach the resonance condition. The main advantage of this resonant electro-optic comb generation over the cascaded EO modulators approach is its highly efficient resonance modulation. It requires $\sim$10 times lower microwave power to generate the same or even larger bandwidth spectra and it requires no external cooling for long-term use.

\begin{figure}[htb]
\centering\includegraphics[width=13.3cm]{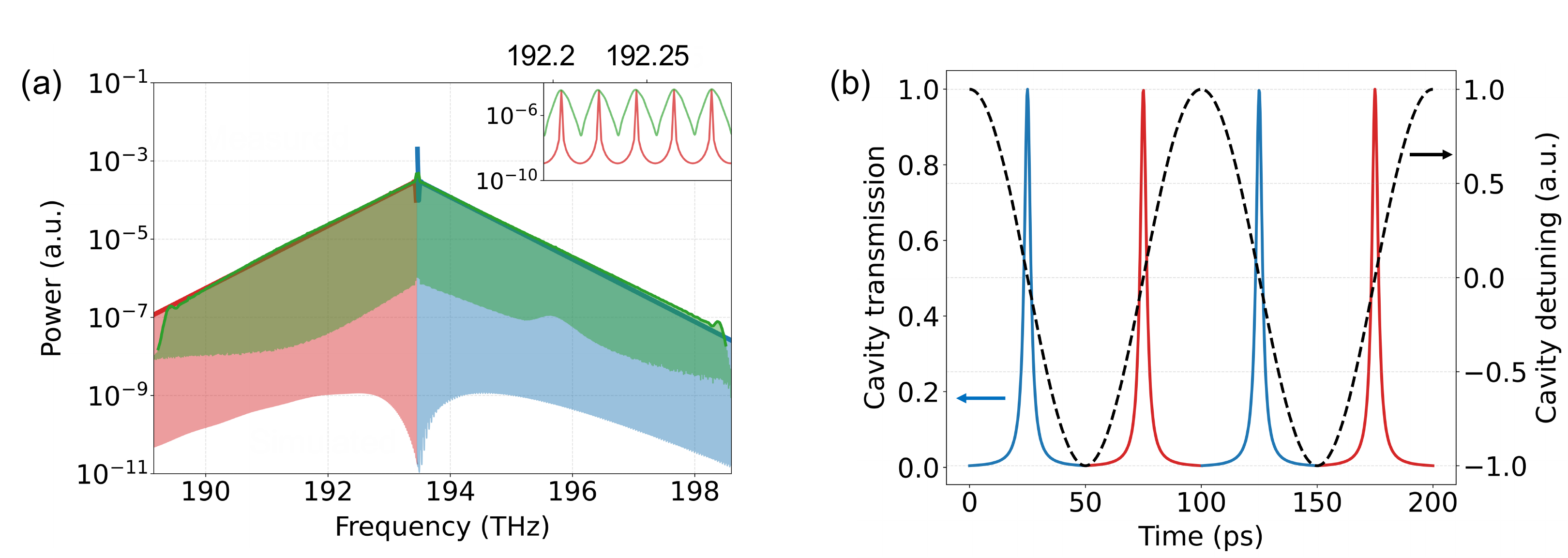}
\caption{(a) Optical spectral output of the REOCG. The measured spectrum is shown by the green trace. The simulated spectrum is given by the red and blue traces. The rise in noise floor due to amplification has not been considered in the simulated spectrum. The inset figure shows the zoomed-in 10 GHz comb modes. (b) Temporal output of the REOCG showing the interleaved pulse trains. Each pulse train corresponds to one side of the spectrum as indicated by the red and blue colors.}
\label{output}
\end{figure}

One of the most interesting features of this resonant comb generator that distinguishes it from cascaded EOM combs is its unique output characteristics in the frequency and time domains \cite{Kobayashi1972High-repetition-rateModulator, Saitoh1998ModulationGenerator, Xiao2008TowardGenerator}. In the frequency domain, the comb envelope resembles a double-sided exponential as shown in Fig. \ref{output}(a). A symmetrical comb spectrum with maximum bandwidth is generated when both the optical and microwave signals are resonant inside the cavity. As the beam circulates forward and backward through a phase modulator inside the cavity, the comb modes above and below the seed frequency develop linear spectral phases with different slopes. This gives rise to a fixed group delay between the two time-domain pulse trains that come into resonance with the cavity as the effective cavity length is modulated (shown as the red and blue Lorentzian traces in Fig. \ref{output}(b)). As a result, the time-domain output of the REOCG consists of two interleaved pulse trains with a relative time delay. This time delay varies from zero to half the modulation period depending on the detuning between the CW laser and the cavity resonance. At resonance, the two interleaved pulse trains appear as a single pulse train at a repetition rate that is twice the modulation frequency applied. In this work, we either overlap the two interleaved pulse trains or remove one of them to achieve higher pulse energy by applying a group delay (in the case of 20 GHz) or by removing half of the optical spectrum (in the case of 10 GHz) using a pulse shaper. When both the seed laser and microwave signal are resonant inside the cavity as in the ideal condition, transmission through the REOCG can be approximated by modifying the transmission formula for a Fabry-P\'erot cavity by including the phase shift from the modulator. Using this assumption, the transmitted electric field from the REOCG in time domain can be written as \cite{Kobayashi1972High-repetition-rateModulator, Saitoh1998ModulationGenerator, Xiao2008TowardGenerator} 
\begin{equation}
E_{t}(t, \beta) \approx \sqrt{\eta} \frac{1-R}{1-R \eta \exp \left[-i \beta \sin \left(2 \pi f_{m} t\right)\right]} E_{i},
\label{eq:refname1}
\end{equation}
where $\eta$ is the single-pass power transmission efficiency of the waveguide modulator ($\eta = 0.976$ \cite{Xiao2008TowardGenerator}), R is the reflectance of the Fabry-P\'erot mirrors (R = 0.97), $\beta = \pi(V/V_{\pi})$ is the modulation index ($\beta \sim 3$ for 1 W RF power) and $f_{m}$ is the modulation frequency. In our device, the total insertion loss is measured to be $\sim$ 25 dB. This includes the fiber-to-waveguide coupling loss as well as the waveguide propagation loss. Interestingly, Loncar \textit{et al.} have demonstrated an improved pump-to-comb conversion efficiency for the resonant EO comb generator on thin-film lithium niobate using two mutually coupled resonators \cite{Hu2022High-efficiencyGenerators}.  

\begin{figure}[htb]
\centering\includegraphics[width=13.3cm]{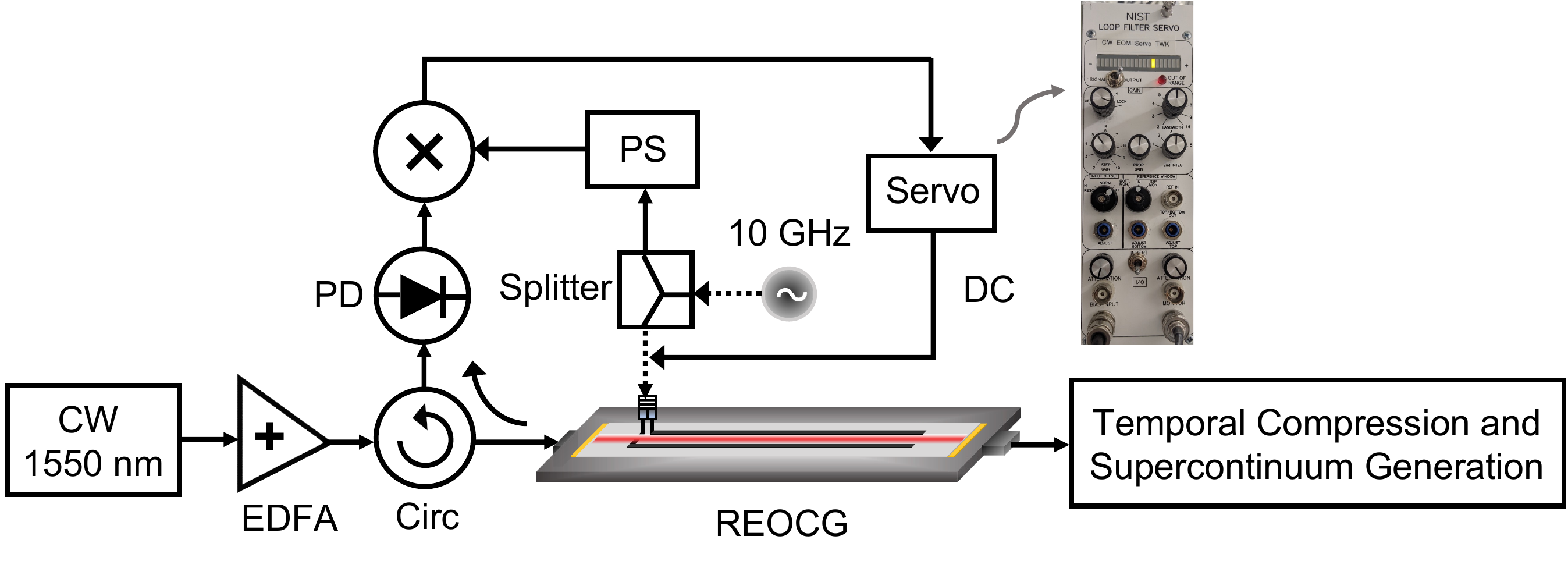}
\caption{The basic layout of the Pound-Drever-Hall technique used for locking the comb generator cavity to the cavity-stabilized continuous wave (CW) laser. EDFA is erbium-doped fiber amplifier, Circ is a fiber circulator, REOCG is the resonant electro-optic comb generator, PD is a photodiode, and PS is a phase shifter.}
\label{lock}
\end{figure}

Fig. \ref{lock} shows the Pound-Drever-Hall (PDH) technique \cite{Drever1983LaserResonator, Black2001AnStabilization} employed for locking the comb generator cavity length so that the detuning between the CW laser and the cavity resonance is held at zero. In this scheme, the CW laser is amplified using an erbium-doped fiber amplifier (EDFA) and then sent to the resonant electro-optic comb generator (REOCG) via a fiber circulator. The light reflected from the comb generator is then collected through one of the circulator ports and measured on a photodetector (PD). The PD output is compared with a phase-shifted modulation signal (10 or 20 GHz) via a mixer. The mixer output signal is then sent to a servo (image shown in Fig. \ref{lock}). The phase of the modulation signal is adjusted to compensate for the unequal delay between the two paths. The proportional and integral gain knobs are then tuned so that the DC output from the servo locks the cavity length to the minimum of the reflected signal. 

\section{Temporal Compression of a Cascaded EOM Comb}

To further show the versatility of our all-fiber temporal compression design, we tried employing the same approach to the output of a frequency comb generated using cascaded electro-optic modulators. In what follows, we describe the details of numerically modeling the comb generator as well as the fiber temporal compression design.

\begin{figure}[htb]
\centering\includegraphics[width=13.3cm]{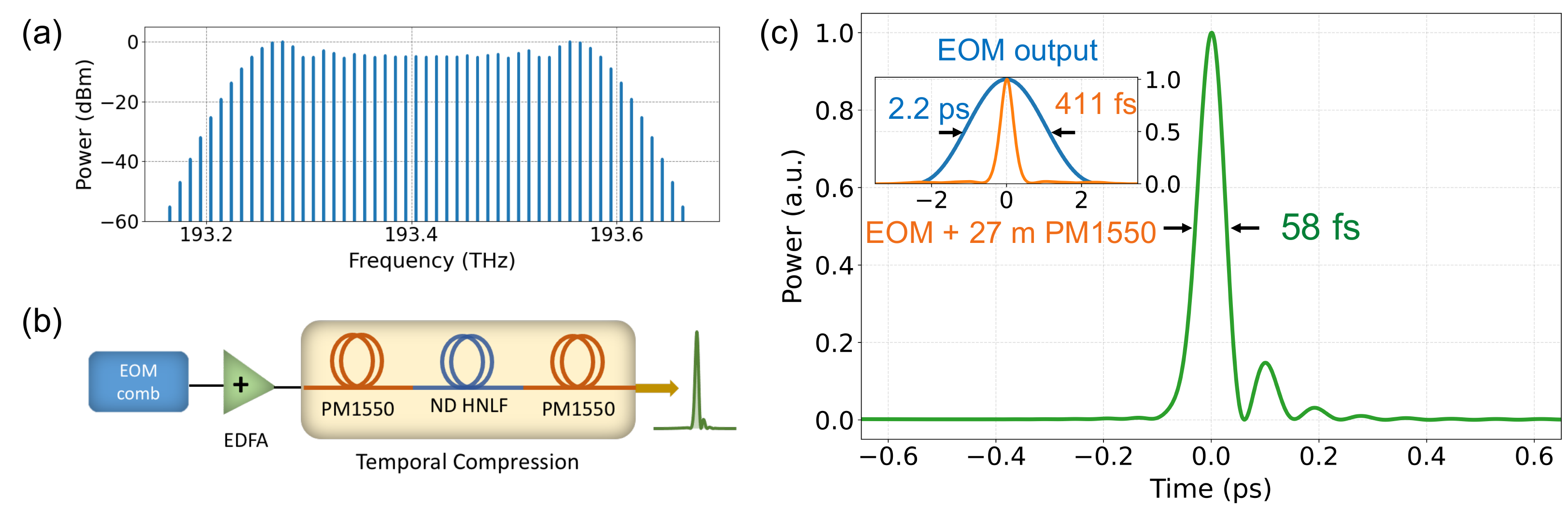}
\caption{(a) Optical spectral output of the cascaded EOM consisting of two phase modulators (PM) and one intensity modulator (IM). (b) Schematic of the all-fiber temporal compressor. (c) A compressed pulse duration of 58 fs is obtained on simulating the propagation of the band-limited EOM output through 27 m PM1550 and then through the temporal compression fiber design consisting of 3.2 m of ND HNLF (D = -2.6 ps/(nm$\cdot$km)) and 56 cm of PM1550. The inset shows the band-limited pulse of the EOM output assuming a flat spectral phase (blue trace) and then propagating it through 27 m of PM1550 for initial temporal compression through soliton effects (orange trace).}
\label{cascaded eom}
\end{figure}
Initially, we simulated a general case in which a CW laser at 1550 nm seeds two phase modulators (PM) and one intensity modulator (IM) arranged in series. Each PM is driven by a 2 W microwave signal at 10 GHz. The resulting simulated 10 GHz comb spanning $\sim$ 0.34 THz (or ~3 nm) is shown in Fig. \ref{cascaded eom}(a). The normal chirp from the phase modulators can be compensated on propagation through PM1550 to obtain a band-limited pulse duration of 2.2 ps (blue trace in the inset of Fig. \ref{cascaded eom}(c)). The above band-limited pulses are then amplified to 3W (pulse energy of 300 pJ at 10 GHz). In order to compress an input pulse duration greater than 1.5 ps with 300 pJ energy using our current all-fiber approach, the simulation (Fig. 1(g) in the main document) predicts more than 36 m of ND HNLF (D = -2.6 ps/(nm. km)) to produce sub-100 fs pulse. %which is not experimentally feasible in terms of optical loss through very long fibers as well as HNLF cost.
It may be possible to shorten the length of ND HNLF required by choosing the HNLF with a larger normal dispersion value at 1550 nm or using additional phase modulators. %However, we are limited by the commercially available options. the required length of ND HNLF might still be beyond the range ($>$ 10 m) that we normally use for these applications in lab. 
However, in order to get around this issue, we simulated propagating the 2.2 ps pulse through PM1550 for an initial temporal compression via soliton self-compression (Fig. \ref{cascaded eom}(b)). This occurs due to the combined effect of anomalous dispersion and self-phase modulation which leads to the formation and propagation of a higher-order soliton. The simulations show that the pulse duration reduces to 411 fs after propagation through 27 m of PM1550 (orange trace in the inset of Fig. \ref{cascaded eom}(c)). 
%This soliton self-compression can be more efficiently done in a shorter length of anomalous dispersion HNLF that has a larger anomalous dispersion value than PM1550 at 1550 nm.

The resultant pulse duration of 411 fs is then considered as input to our temporal compression fiber design. Further compression can be efficiently achieved by spectral broadening in ND HNLF and chirp compensation in PM1550. The simulation results give a compressed pulse duration of 58 fs on propagating through 3.2 m of ND HNLF and 56 cm of PM1550 as shown in Fig. \ref{cascaded eom}(c). Using this all-fiber design, we are able to predict a compression factor of $\sim$ 38 which is even greater than that achieved in nonlinear nanophotonic waveguides \cite{Oliver2021Soliton-effectChip}.

\begin{figure}[h!]
\centering\includegraphics[width=13.3cm]{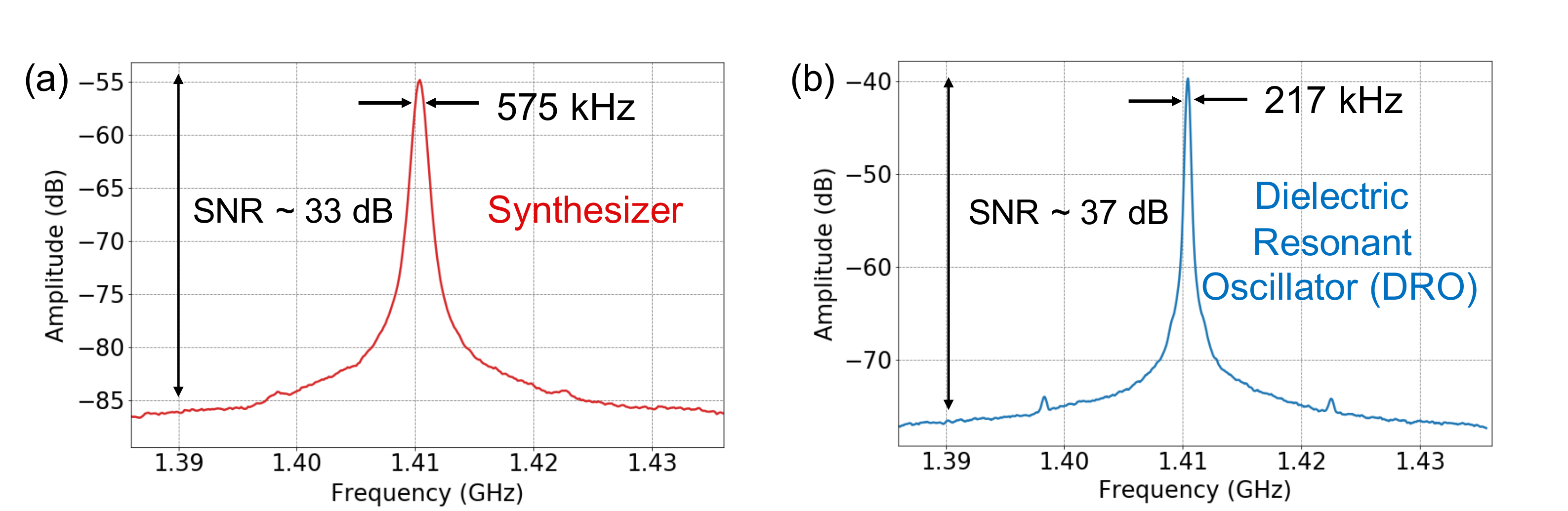}
\caption{Heterodyne beat between the optically filtered supercontinuum and a 1319 nm CW laser with (a) an Agilent synthesizer, and (b) a DRO as source of the initial 20 GHz signal. The linewidth and SNR of the beat signals are noted.}
\label{het}
\end{figure}
\section{Supercontinuum coherence}

We performed optical heterodyne measurements as an initial step to check the coherence of the 20 GHz supercontinuum obtained with multi-segment AD HNLF. Initially, the supercontinuum is passed through a bandpass filter centered at 1319 nm with a bandwidth of 12 nm. The heterodyne beat between the filtered 20 GHz comb modes and a stable 1319 nm CW laser (JDSU model: M126N-1319-350) on a photodiode is amplified and observed on the radio frequency spectrum analyzer (RFSA) as shown in Fig. \ref{het}. We also studied the impact that the choice of the RF oscillator had on the phase noise of the comb modes. The signal-to-noise ratio of the heterodyne beat increased from $\sim$ 33 dB (Fig. \ref{het}(a)) with an Agilent E8257N synthesizer to 37 dB  using a dielectric resonant oscillator (DRO) as RF source (Fig. \ref{het}(b)). Both signals were measured on an RFSA with a resolution bandwidth (RBW) of 100 kHz. It is also observed that the 3 dB linewidth of the beat signal reduces by more than a factor of two on using the DRO as an RF source instead of a synthesizer. These results indicate that we can improve the coherence of our supercontinuum to a large extent by using low-noise RF oscillators and phase-locked loops. In the future, we plan to extend this heterodyne beat measurement to frequencies further away from the pump and study in detail the impact of input amplitude noise, and of the built-in cavity's filtering of the broadband microwave thermal noise, on the coherence of the generated supercontinuum. 
%The measured high SNR shows that the all-fiber technique in this work provides a robust way to generate a low-noise coherent broadband supercontinuum from off-the-shelf components.

\bibliography{osa-supplemental-document-template}